\documentclass[aps,prl,twocolumn,superscriptaddress,showpacs,floatfix]{revtex4}

\usepackage{graphicx}
\usepackage{dcolumn}
\usepackage{bm}

\begin{document}


\title{Absolute Branching Fraction Measurements for\\
         $D^+$ and $D^0$ Inclusive Semileptonic Decays}

\author{N.~E.~Adam}
\author{J.~P.~Alexander}
\author{K.~Berkelman}
\author{D.~G.~Cassel}
\author{J.~E.~Duboscq}
\author{K.~M.~Ecklund}
\author{R.~Ehrlich}
\author{L.~Fields}
\author{L.~Gibbons}
\author{R.~Gray}
\author{S.~W.~Gray}
\author{D.~L.~Hartill}
\author{B.~K.~Heltsley}
\author{D.~Hertz}
\author{C.~D.~Jones}
\author{J.~Kandaswamy}
\author{D.~L.~Kreinick}
\author{V.~E.~Kuznetsov}
\author{H.~Mahlke-Kr\"uger}
\author{T.~O.~Meyer}
\author{P.~U.~E.~Onyisi}
\author{J.~R.~Patterson}
\author{D.~Peterson}
\author{J.~Pivarski}
\author{D.~Riley}
\author{A.~Ryd}
\author{A.~J.~Sadoff}
\author{H.~Schwarthoff}
\author{X.~Shi}
\author{S.~Stroiney}
\author{W.~M.~Sun}
\author{T.~Wilksen}
\author{M.~Weinberger}
\affiliation{Cornell University, Ithaca, New York 14853}
% ...
\collaboration{CLEO Collaboration} 
\noaffiliation

\date{November 13, 2006}

\begin{abstract}
We present measurements of the inclusive branching fractions for the
decays $D^+\rightarrow X e^+ \nu _e$ and $D^0\rightarrow X e^+ \nu
_e$, using 281 pb$^{-1}$ of data collected on the $\psi(3770)$
resonance with the CLEO-c detector. We find ${\cal B}(D^0\rightarrow
Xe^+\nu_e) = (6.46 \pm 0.17 \pm 0.13)\%$ and ${\cal
B}(D^+\rightarrow Xe^+\nu_e) = (16.13 \pm 0.20 \pm 0.33)\%$. Using
the known $D$ meson lifetimes, we obtain the ratio
$\Gamma_{D^+}^{\text{sl}}/\Gamma_{D^0}^{\text{sl}}= 0.985\pm
0.028\pm 0.015$, confirming isospin invariance at the level of 3\%.
The positron momentum spectra from  $D^+$ and $D^0$ have consistent
shapes.
\end{abstract}

 \maketitle

The study of inclusive $D$ semileptonic decays is important for
several reasons. First, by comparing the inclusive branching
fractions of the $D^+$ and $D^0$ mesons  with the sum of the
measured exclusive branching fractions, one can determine whether
there are unobserved semileptonic decay modes. Previous data suggest
that the lightest vector and pseudoscalar resonances saturate the
hadronic spectra \cite{pdg04}. This may be due to the relatively low
momentum of the daughter $s$ quark, that favors the formation of
s-wave hadrons. Alternatively, this may be an indication that heavy
quark effective theory may still be valid at the charm quark mass
scale \cite{isgw2}. In addition, since accurate experimental
determinations of the $D^0$ and $D^+$ lifetimes are available
\cite{pdg04}, measurements of semileptonic branching fractions
determine the corresponding semileptonic widths,
$\Gamma_{D^+}^{\text{sl}}$ and $\Gamma_{D^0}^{\text{sl}}$. These
widths are expected to be equal, modulo small corrections introduced
by electromagnetic effects. Weak annihilation diagrams can produce
more dramatic effects on the Cabibbo suppressed partial widths
\cite{bigi-bianco}. As these contributions may also influence
 the extraction of $V_{ub}$ from inclusive $B$ meson
semileptonic decays, it is important to understand them well.
Finally, better knowledge of the inclusive positron spectra can be
used to improved modeling of the ``cascade'' decays $b\rightarrow
c\rightarrow s e^+ \nu _e$ and thus is important in several
measurements of $b$ decays.

  The use of ratios of semileptonic branching fractions as a probe
of relative lifetimes of the $D$ mesons was suggested by Pais and
Treiman \cite{pt}. Indeed the early measurements of the ratio of the
$D^+$ and $D^0$ semileptonic branching fractions gave the first
surprising evidence for the lifetime difference between these two
charmed mesons \cite{schindler, bacino}. Later, the first
measurement of the individual charged and neutral $D$ inclusive
semileptonic branching fractions was performed by Mark III
\cite{baltrusaitis}, with an overall relative error of about 12-16\%
on the individual branching fractions, and 19\% on their ratio. The
inclusive decay $D^0 \rightarrow X e^+ \nu _e$ was subsequently
studied by ARGUS \cite{argus-d0} and CLEO \cite{cleo-d0}, using the
angular correlation between the $\pi ^+$ emitted in a $D^{\star
+}\rightarrow \pi^+ D^0$ decay and the $e^+$ emitted in the
subsequent $D^0\rightarrow X e^+ \nu _e$ decay. The more precise
CLEO result has a 5\% relative error, dominated by systematic
uncertainties. Our measurements exploit a clean $D\bar{D}$ sample at
threshold, and thus achieves significantly smaller systematic
errors, through low backgrounds and well understood efficiencies.

We use a 281 pb$^{-1}$ data sample, collected at the $\psi({\rm
3770})$ center-of-mass energy ($\sqrt{s}\approx$ 3.73 GeV), with the
CLEO-c detector \cite{cleoiii}. This detector includes a tracking
system composed of a six-layer low-mass drift chamber and a 47-layer
central drift chamber, measuring charged particle momentum and
direction, a state of the art CsI(Tl) electromagnetic calorimeter,
and a Ring Imaging Cherenkov (RICH) hadron identification system.
All these components are critical to an efficient and highly
selective electron and positron identification algorithm. The
charged particle momentum resolution is approximately 0.6\% at 1
GeV. The CsI(Tl) calorimeter measures the electron and photon
energies with a resolution of 2.2\% at $E=1$ GeV and 5\% at $E$=100
MeV, which, combined with the excellent tracking system, provides
one of the $e$ identification variables, $E/p$, where $E$ is the
energy measured in the calorimeter and $p$ is the momentum measured
in the tracking system. The tracking system provides charged
particle discrimination too, through the measurement of the specific
ionization $dE/dx$. Charged particles are also identified over most
of their momentum range in the RICH detector \cite{rich}. In
particular, RICH identification plays a crucial role at momenta
where the specific ionization bands of two particle species cross
each other and $dE/dx$ does not provide any discrimination power.


We use a tagging technique similar to the one pioneered by the Mark
III collaboration \cite{markIII}. Details on the tagging selection
procedure are given in Ref.~\cite{dhad}. We select events containing
either the decay $\bar{D}^0 \rightarrow K^+\pi^-$ or the decay
$D^-\rightarrow K^+\pi ^-\pi^-$. We use only these modes, because
they have very low background. Note that charge conjugate modes are
implied throughout this paper. In this analysis we exploit the
flavor information provided by the tagging $D$:  the $D^-$ charge
sign provides a flavor tag, whereas the charge of the tag daughter
$K$ is used for $\bar{D}^0$ flavor assignment.

We analyze all the recorded events at the $\psi(3770)$ and retain
the events that contain at least one candidate $\bar{D}^0\rightarrow
K^+\pi ^-$
 or $D^-\rightarrow K^+\pi ^-\pi ^-$. Two kinematic variables are
 used to select these candidates: the
beam-constrained mass, $M_{\rm bc}\equiv \sqrt{E_{\rm beam}^2-
(\Sigma _i \vec{p}_i)^2}$, and the energy difference $\Delta E$,
where $\Delta E\equiv ( \Sigma _i E_i- E_{\rm beam})$, where $E_{\rm
beam}$ represents the beam energy and $(E_i,\vec{p}_i)$ represent
the 4-vectors of the candidate daughters. For $\bar{D}^0$ tags, the
measured standard deviations ($\sigma$) in $\Delta E$ and $M_{\rm
bc}$ are $\sigma(\Delta E)= 6.6$ MeV and $\sigma(M_{\rm bc}) = 1.35$
MeV, while for ${D}^-$ tags, $\sigma(\Delta E)= 5.9$ MeV and
$\sigma(M_{\rm bc}) = 1.34$ MeV. We select events that are within 3
$\sigma$ of the expected $\Delta E$ (0 GeV) and $M_{\rm bc}$ ($M_D$)
for the channels considered. In order to determine the total number
of tags, we count the events within the selected $\Delta E$-$M_{\rm
bc}$ intervals; then we subtract the combinatoric background
inferred from two $3\sigma$ sideband regions on both sides of the
$\Delta E=0$ signal peak, with a $2\sigma$ gap from the signal
interval. The yields in the signal region are $48204\ \bar{D}^0$ and
$76635\ D^-$. The corresponding yields in the sideband region are
$788\pm 28$ and $2360\pm 49$. We correct the sideband yields with
scale factors accounting for the relative area of the background in
the signal and sideband intervals (1.047 for $\bar{D}^0$ and 1.23
for $D^-$). The scaling factors are inferred from the background
component of the $M_{\rm bc}$ fits. We obtain $47379\pm 29
\bar{D}^0$ tagged events, and $73732\pm 60\ D^-$ tagged events. As
we are  interested in counting the number of signal events, and not
in measuring a production rate, the errors only reflect the
uncertainty in the background subtraction. Note that the estimated
background is only 1.7\% of the signal for $\bar{D}^0$ and 3.9\% for
$D^-$.


For each event selected, we study all the charged tracks not used in
the tagging mode. We select the ones that are well-measured, and
whose helical trajectories approach the event origin within a
distance of 5 mm in the projection transverse to the beam and 5 cm
in the projection along the beam axis. Each track must include at
least 50\% of the hits expected for its momentum. Moreover, it must
be within the RICH fiducial volume
 ($|\cos(\theta)| \le 0.8$), where $\theta$ is the angle with respect to the
 beams. Finally,
we require the charged track momentum $p_{\rm track}$ to be greater
than or equal to 0.2 GeV, as the particle species separation becomes
increasingly difficult at low momenta.

 Candidate positrons (and electrons) are
selected on the basis of a likelihood ratio constructed from three
inputs: the ratio between the energy deposited in the calorimeter
and the momentum measured in the tracking system, the specific
ionization $dE/dx$ measured in the drift chamber, and RICH
information \cite{d0excl}. Our particle identification selection
criteria have an average efficiency of 0.95 in the momentum region
0.3-1.0 GeV, and 0.71 in the region 0.2-0.3 GeV.

The $e^+$ sample contains a small fraction of hadrons that pass our
selection criteria. As the probability that a $\pi$ is identified
 as an $e$  at a given momentum is different from the
corresponding $K$ to $e$ misidentification probability, we need to
know
 the $K$ and $\pi$  yields separately to subtract this background.
 We select $\pi$ and $K$ samples using a particle identification
variable (PID) that combines RICH and $dE/dx$ information, if the
RICH identification variable \cite{rich} is available and $p_{\rm
track}>0.7\ {\rm GeV}$; alternatively PID relies on $dE/dx$ only.
The $\pi$ sample contains also  a $\mu$ component, as our PID
variable is not very selective; however, as our goal is only to
unfold the true $e$ spectrum, we do not need to correct for this
effect.

We separate $e$, $\pi$, and $K$ into ``right-sign'' and
``wrong-sign'' samples according to their charge correlation to the
flavor tag. Right-sign assignment is based on the expected $e$
charge on the basis of the flavor of the decaying $D$. The true $e$
populations in the right-sign and wrong-sign samples are obtained
through an unfolding procedure, using the matrix:
\[ \left( \begin{array}{c}     n_{e}^{m} \\
     n_{\pi}^{m} \\
     n_{K}^{m}
    \end{array} \right)  =
\left( \begin{array}{ccc}
  \varepsilon_{e} & f_{e \pi} & f_{eK}   \\
  f_{\pi e} & \varepsilon_{\pi} & f_{ \pi K}   \\
  f_{Ke} &     f_{K \pi} & \varepsilon_{K}
\end{array} \right)
\times \left( \begin{array}{c}
n_{e}^{t}\\
(n_{\pi}^{t} +\kappa  n_{\mu}^t)\\
n_K^{t}
\end{array} \right);\]
here $n_{e}^m$, $n_{\pi}^m$, $n_{K}^m$ represent the raw measured
spectra in the corresponding particle species, and the coefficient
$\kappa$ accounts for the fact that the efficiencies for $\pi$ and
$\mu$ selection are not necessarily identical, especially at low
momenta. The quantities $n_{e}^t$, $n_{\pi}^t$, and $n_{K}^t$
represent the true $e$, $\pi$ and $K$ spectra: the present paper
focuses on the extraction of $n_{e}^t$. As the $\pi$ to $e$
misidentification probability is quite small, the effect of a small
$\mu$ component in the measured $\pi$ population $n_{\pi}^m$ is
negligible.
 The efficiencies $\epsilon _e,\ \epsilon _{\pi},\ {\rm
and}\ \epsilon _K$ account for track finding, track selection
criteria, and particle identification losses. The tracking
efficiencies are obtained from a Monte Carlo simulation of
$D\bar{D}$ events in the CLEO-c detector. The generator incorporates
all the known $D$ decay properties, includes initial state radiation
(ISR), and final state radiation (FSR) effects, the latter are
modeled with the program PHOTOS \cite{photos}. The particle
identification efficiencies are determined from data. We study the
$K$ selection efficiencies using a sample of $D^+\rightarrow
K^-\pi^+\pi^+$ decays, and $\pi$ selection efficiencies using
$D^+\rightarrow K^-\pi^+\pi^+$ and $K^0_S\rightarrow \pi^+\pi^-$
decays. The $e^+$ identification efficiency is extracted from a
radiative Bhabha sample. A correction for the difference between the
$D\bar{D}$ event environment and the simpler radiative Bhabha
environment (two charged tracks and one shower) is derived using a
Monte Carlo sample where a real electron track stripped from a
radiative Bhabha event is merged with tracks from a simulated
hadronic environment. The off-diagonal elements are products of
tracking efficiencies and particle misidentification probabilities,
where $f_{ab}$ is defined as the probability that particle $b$ is
identified as particle $a$. The $f_{ab}$ parameters are determined
using $e$ samples from radiative Bhabhas, and $K$ and $\pi$ from
$D^+ \rightarrow K^- \pi^+\pi^+$ and $K^0_S\rightarrow \pi^+\pi^-$.
The $e$ spectrum from radiative Bhabhas is divided in 50 MeV
momentum bins to determine the corresponding misidentification
probabilities, whereas the $K$ and $\pi$ populations are subdivided
into 100     MeV momentum bins  to reduce the statistical
uncertainty. The hadron to $e^+$ misidentification probabilities are
of the order of 0.1\% over most of the momentum range and below 1\%
even in the regions where $dE/dx$ separation is less effective.

There are background sources that are charge symmetric, mostly
produced by $\pi ^0$ Dalitz decays and $\gamma$ conversions. We
subtract the wrong sign unfolded yields from the corresponding right
sign yields to account for them, motivated by Monte Carlo studies
that confirm the accuracy of this method.

 In order to subtract the combinatoric
background, we repeat the unfolding procedure determining the true
$e^+$ yields from measured $e^+$, $\pi^+$, and $K^+$ samples where
the tags are selected from $\Delta E$ sidebands. The decays
$\bar{D}^0\rightarrow K^+\pi^-$ and $D^-\rightarrow K^+\pi^-\pi^-$
have very little background: the $\bar{D}^0$ sidebands give a
combinatoric background estimate that is 0.2\% of the signal yield
and the $D^-$ sidebands give a combinatoric background estimate that
is 1.8\% of the signal yield. Table \ref{tab:cleantags} shows the
results of the intermediate steps involved in the determination of
the net $e^+$ yields. Efficiency corrections increase unfolded $e^+$
yields with respect to uncorrected $e^+$ yields, while the
subtraction of the contribution from misidentified hadrons reduces
them. The former effect is dominant for the right-sign positron
sample, while it is comparable in size to the background subtraction
in the wrong-sign sample. The final yields, identified as
``corrected net $e^+$'' include acceptance corrections related to
the ($\cos{\theta} \le 0.8$) cut, and doubly-Cabibbo suppressed
(DCSD) effects in $D^0$ decays. As we are using the charge of the
tagging $D^-$, rather than its $K$ daughter charge, this correction
is not needed in the charged mode.

\begin{table}[htpb]
\begin{center}
\caption{\label{tab:cleantags}  Positron unfolding procedure and
corrections. The errors reported in the intermediate yields reflect
only statistical uncertainties.} \vskip 0.5 cm
\begin{tabular}{lcc}
\hline ~~& $D^+$ & $D^0$ \\
\hline
Signal $e^+$& ~~ & ~~ \\
Right-sign & $8275\pm 91$ & $ 2239\pm 47$ \\
Wrong-sign  & $228\pm 15$ & $233 \pm 15$ \\
\hline Right-sign (unfolded) & $9186\pm 103$ & $2453\pm 54$\\
Wrong-sign (unfolded) & $231\pm 19$ & $ 203\pm 19$ \\
Sideband $e^+$(RS) & 168$\pm$ 13 & 15 $\pm$ 4 \\
Sideband $e^+$(WS) & $11\pm 5$ & $11\pm 4$ \\
Net $e^+$ & $8798 \pm 105$ & $2246\pm 57$ \\
Corrected Net $e^+$ & $10998\pm 132$ & $2827\pm 72$\\\hline
\end{tabular}
\end{center}
\end{table}

In order to extract the partial branching fractions for $p_e \ge
0.2$ GeV, we evaluate the ratio between the net positron yields
corrected for geometric acceptance and the net number of tags. While
the charge of $D^-\rightarrow K^+\pi^-\pi^-$ reliably tags the
flavor of the charged $D$, in the $\bar{D}^0$ case the $K$ charge
occasionally produces an incorrect flavor assignment due to the DCSD
$\bar{D}^0\rightarrow K^-\pi^+$. This effect is estimated on the
basis of the known value of the parameter $r_{\text{DCSD}}\equiv
N(D^0\rightarrow K^+\pi^-)/N(D^0\rightarrow K^-\pi^+)= 0.00362\pm
0.00029$ \cite{pdg04}.

We have considered several sources of systematic uncertainties.
There are multiplicative errors that affect the overall scale of the
spectrum, including tracking efficiency or electron identification
efficiency, accounting for Monte Carlo modeling uncertainties. The
uncertainties in tracking and $K$ and $\pi$ identification
efficiencies are taken from the studies discussed in
Ref.~\cite{dhad}. The systematic error on the electron
identification efficiency (1\%) is assessed by comparing radiative
Bhabha samples, radiative Bhabha tracks embedded in
 $D\bar{D}$ Monte Carlo samples, and $D\bar{D}$ Monte Carlo samples. These
 contributions are common to $D^+$
and $D^0$. In addition, we have accounted for the FSR uncertainty by
varying its amount, with a total systematic error of 0.5\%. The last
multiplicative error is the uncertainty on the number of tags,
estimated by comparing the number of background tags in our signal
window from the $\Delta E$ sidebands and from $M_{\text{bc}}$
sidebands. In addition, there are terms that are affected by limited
statistics, such as misidentification probabilities, or particle
identification efficiencies. The systematic uncertainty associated
with these terms is evaluated with a toy Monte Carlo; we perform
$10^6$ iterations of the unfolding procedure,  and vary the matrix
elements within error. The corresponding  relative systematic error
estimates are 0.56\% (statistical errors on particle
misidentification probability and particle identification
efficiency) and 0.3\% (statistical error on tracking efficiency).
The uncertainty on the combinatoric background, accounted for with
the sideband positron sample, is negligible compared with these
components ($\le 0.1\%)$, because of the excellent purity of the tag
samples used. Thus the total relative systematic error on the
branching fraction for $p_e\ge 0.2$\ GeV is 1.7\% ($D^0$) and 1.8 \%
($D^+$).

The partial branching fractions for $p_e \ge 0.2$ GeV are evaluated
as the ratio between the corrected net $e$ yields and the net number
of tags:
   $${\cal B}(D^{+}\rightarrow Xe^+\nu _e) = (14.92 \pm 0.19_{\rm stat} \pm
   0.27_{\rm sys})\%;$$
   $${\cal B}(D^{0}\rightarrow Xe^+\nu _e)=  (5.97 \pm 0.15_{\rm stat} \pm
   0.10_{\rm sys})\%.$$

The yield in the unmeasured region ($p_e <0.2$ GeV) is estimated by
fitting the measured spectra with a shape derived from Monte Carlo.
The semileptonic decays are generated with the ISGW form factor
model \cite{isgw}, with parameters tuned to experimental constraints
such as measured branching fractions, with the procedure described
in Ref.~\cite{cleo-d0}. Final state radiation effects are included
in the simulation. We obtain $f(p_e)\equiv \Delta \Gamma_
{\text{sl}} (p_e < 0.2\ {\rm GeV})/\Gamma_{\text{sl}}= (7.5\pm
0.5)\%$ for $D^+\rightarrow X e^+ \nu _e$ and $\Delta
\Gamma_{\text{sl}}(p_e < 0.2\ {\rm GeV})/\Gamma_{\text{sl}}= (7.7\pm
0.9)\%$ for $D^0\rightarrow Xe^+ \nu _e$. The $\chi ^2$ per degree
of freedom is 1.23 for $D^+\to X e^+\nu _e$, 0.75 for $D^+\to X
e^+\nu _e$. We studied the sensitivity of our analysis to $f(p_e)$
using alternative fitting procedures, such as a combination of the
dominant exclusive channels modeled with different form factors
\cite{isgw2}. The fractional difference in $f(p_e)$ with the various
methods considered is below 4\% and is well within the systematic
errors assigned. Note that the relative error in the branching
fraction introduced by the extrapolation to the unmeasured portion
of the spectrum is given $\delta f(p_e)/(1-f(p_e))$, and thus the
systematic error on the total semileptonic branching fractions is a
about 1\%.
 Upon applying this correction, we obtain:
   $${\cal B}(D^{+}\rightarrow Xe^+\nu _e) = (16.13 \pm 0.20_{\rm stat} \pm
   0.33_{\rm sys})\%;$$
   $${\cal B}(D^{0}\rightarrow Xe^+\nu _e) =  (6.46 \pm 0.17_{\rm stat}  \pm
   0.13_{\rm sys})\%.$$

\begin{figure}[htbp]
\includegraphics*[width=3.5in]{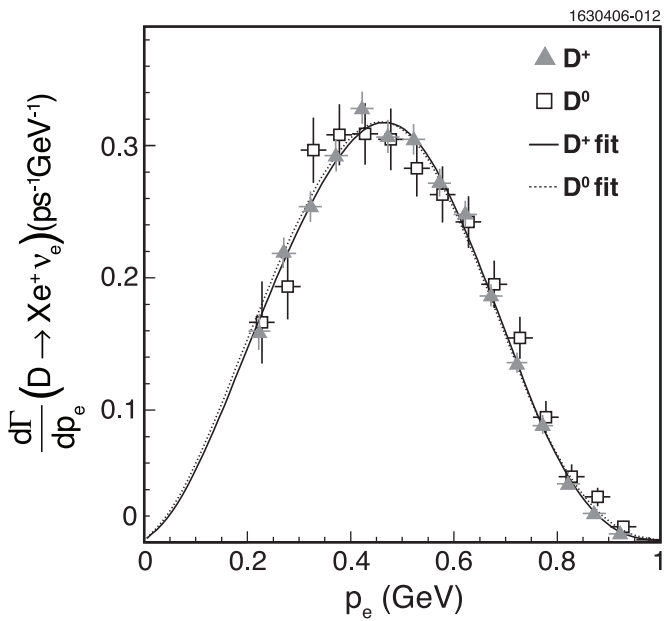} \caption{\small Positron differential
semileptonic widths $d\Gamma^{\text{sl}}/dp_e$ for  the decays
$D^0\rightarrow X e^+ \nu_e$ (open squares) and $D^+\rightarrow Xe^+
\nu_e$ (filled triangles) in the laboratory frame. The errors shown
include statistical and additive systematic errors. The symbols for
$D^+$ and $D^0$ spectra are slightly shifted horizontally to avoid
overlapping. The curves are derived from the fits used to
extrapolate the measured spectra below the $p^{\text{min}}$
cut.}\label{fig:labspectra}
\end{figure}

Using the well-measured lifetimes of the $D^+$ and $D^0$ mesons,
$\tau_{D^+}=(1.040\pm 0.007)$ ps, and $\tau_{D^0}=(0.4103\pm
0.0015)$ ps \cite{pdg04},  we normalize the measured partial
branching fractions to obtain differential semileptonic widths.
Figure \ref{fig:labspectra} shows the differential semileptonic
widths $d\Gamma^{\text{sl}}/dp_e$ in the laboratory frame, where the
$D^+$ momentum is 0.243 GeV, and the $D^0$ momentum is 0.277 GeV. No
final state radiation correction is applied to the data points.
Table~\ref{tab:labspectra} shows the corresponding numerical values.
The errors shown are evaluated by adding the statistical errors and
the additive systematic errors in quadrature. In addition, an
overall multiplicative systematic error of about 1.5\% needs to be
included in derived quantities such as the total semileptonic width
to account for overall tracking and particle identification
efficiency uncertainties. The total inclusive semileptonic widths
are $\Gamma(D^{+}\rightarrow Xe^+\nu _e) = 0.1551 \pm 0.0020 \pm
0.0031$ ps$^{-1}$, and $\Gamma(D^{0}\rightarrow Xe^+\nu _e) = 0.1574
\pm 0.0041 \pm 0.0032$ ps$^{-1}$. The corresponding ratio of the
semileptonic widths of charged and neutral $D$ mesons is
$\Gamma_{D^+}^{\text{sl}}/\Gamma_{D^0}^{\text{sl}}=0.985\pm 0.028\pm
0.015$, consistent with isospin invariance.

\begin{table}[htpb]
\begin{center}
\caption{ $D^+$  and $D^0$ positron differential semileptonic widths
$d\Gamma/dp_e$(ps$^{-1}$GeV$^{-1}$) in the laboratory frame. The
errors shown include statistical errors and additive systematic
errors \label{tab:labspectra}.}
\begin{tabular}{lcc}
\hline
 $p_e$ (GeV) & $d\Gamma/dp_e(D^+)$ & $d\Gamma/dp_e(D^0)$ \\ \hline
 0.20 - 0.25 & 0.1598 $\pm$ 0.0142 & 0.1664 $\pm$ 0.0311\\
 0.25 - 0.30 & 0.2185 $\pm$ 0.0121 & 0.1935 $\pm$ 0.0248\\
 0.30 - 0.35 & 0.2538 $\pm$ 0.0116 & 0.2966 $\pm$ 0.0247\\
 0.35 - 0.40 & 0.2925 $\pm$ 0.0121 & 0.3081 $\pm$ 0.0231\\
 0.40 - 0.45 & 0.3281 $\pm$ 0.0127 & 0.3088 $\pm$ 0.0233\\
 0.45 - 0.50 & 0.3064 $\pm$ 0.0130 & 0.3047 $\pm$ 0.0233\\
 0.50 - 0.55 & 0.3047 $\pm$ 0.0115 & 0.2828 $\pm$ 0.0214\\
 0.55 - 0.60 & 0.2716 $\pm$ 0.0111 & 0.2631 $\pm$ 0.0212\\
 0.60 - 0.65 & 0.2479 $\pm$ 0.0104 & 0.2422 $\pm$ 0.0196\\
 0.65 - 0.70 & 0.1864 $\pm$ 0.0088 & 0.1951 $\pm$ 0.0179\\
 0.70 - 0.75 & 0.1359 $\pm$ 0.0076 & 0.1547 $\pm$ 0.0158\\
 0.75 - 0.80 & 0.0892 $\pm$ 0.0060 & 0.0948 $\pm$ 0.0121\\
 0.80 - 0.85 & 0.0444 $\pm$ 0.0042 & 0.0498 $\pm$ 0.0091\\
 0.85 - 0.90 & 0.0221 $\pm$ 0.0028 & 0.0344 $\pm$ 0.0070\\
 0.90 - 0.95 & 0.0065 $\pm$ 0.0015 & 0.0120 $\pm$ 0.0044\\
 0.95 - 1.00 & 0.0007 $\pm$ 0.0005 & 0.0020 $\pm$ 0.0020\\
\hline
\end{tabular}
\end{center}
\end{table}

Finally, we can compare these widths with the sum of the
semileptonic decay widths for the pseudoscalar and vector hadronic
final states recently published by CLEO \cite{cleo:dplus}:  ${\cal
B}(D^{+}\rightarrow Xe^+\nu _e)_{\rm excl}= (15.1 \pm 0.5 \pm
0.5)$\% and ${\cal B}(D^{0}\rightarrow Xe^+\nu _e)_{\rm excl}= (6.1
\pm 0.2 \pm 0.2)$\%: the measured
 exclusive modes are consistent with saturating the
inclusive widths, although there is some room left for higher
multiplicity modes. The composition of the inclusive hadronic
spectra is dominated by the low lying resonances in the
$c\rightarrow s$ and $c\rightarrow d$, in striking contrast with $B$
semileptonic decays, where a sizeable component of the inclusive
branching fraction is still unaccounted for \cite{pdg04}.

In conclusion, we report improved measurements of the absolute
branching fractions for the inclusive semileptonic decays ${\cal
B}(D^+\rightarrow X e^+ \nu _e)=(16.13\pm 0.20\pm 0.33)\%$ and
${\cal B}(D^0\rightarrow X e^+ \nu _e)=(6.46\pm 0.17\pm 0.13)\%$.
Using the measured $D$ meson lifetimes, the ratio
$\Gamma_{D^+}^{\text{sl}}/\Gamma_{D^0}^{\text{sl}}=0.985\pm 0.028\pm
0.015$ is extracted, and it is consistent with isospin invariance.
The shapes of the spectra are consistent with one another within
error.

\section{Acknowledgements}
We gratefully acknowledge the effort of the CESR staff in providing
us with excellent luminosity and running conditions. This work was
supported by the A.P.~Sloan Foundation, the National Science
Foundation, the U.S. Department of Energy, and the Natural Sciences
and Engineering Research Council of Canada.


\begin{thebibliography}{00}
\bibitem{pdg04}
S.~Eidelman {\it et al.}, Phys.\ Lett.\ B {\bf 592}, 1 (2004).
\bibitem{isgw2}
D.~Scora and N.~Isgur,
  Phys.\ Rev.\ D {\bf 52}, 2783 (1995)
  [arXiv:hep-ph/9503486].
\bibitem{bigi-bianco}
S.~Bianco, F.~L.~Fabbri, D.~Benson and I.~Bigi,
Riv.\ Nuovo Cim.\  {\bf 26N7}, 1 (2003) [arXiv:hep-ex/0309021].
\bibitem{pt}A.~Pais and S.~B.~Treiman, Phys.\ Rev.\ D {\bf 15}, 2529 (1977).
\bibitem{schindler} R.~H.~Schindler {\it et al.} [Mark II
Collaboration], Phys. Rev. D {\bf 24}, 78 (1981).
\bibitem{bacino} W. Bacino {\it et al.} [DELCO Collaboration], Phys.\ Rev.\
Lett.\ {\bf 45}, 329 (1980).
\bibitem{baltrusaitis}
R.~M.~Baltrusaitis {\it et al.}  [Mark III Collaboration],
  Phys.\ Rev.\ Lett.\  {\bf 54}, 1976 (1985)
  [Erratum-ibid.\  {\bf 55}, 638 (1985)].
\bibitem{argus-d0}
H. Albrecht {\it et al.}  [ARGUS Collaboration], Phys.\ Lett.\ B
{\bf 374}, 249 (1996).

\bibitem{cleo-d0}
Y.~Kubota {\it et al.}  [CLEO Collaboration], \ D {\bf 54}, 2994
(1996).

\bibitem{cleoiii}
Y. Kubota {\it et al.}, Nucl.\ Instrum.\ Meth.\ A {\bf 320}, 66
(1992).
\bibitem{rich}
M.~Artuso {\it et al.},
Nucl.\ Instrum.\ Meth.\ A {\bf 502}, 91 (2003)
[arXiv:hep-ex/0209009].

\bibitem{markIII}J. Adler {\em et al.} [Mark III Collaboration],
Phys. Rev. Lett. {\bf 62}, 1821 (1989).
\bibitem{dhad}Q. He {\em et al.} [CLEO Collaboration],
Phys.\ Rev.\ Lett.\  {\bf 95}, 121801 (2005) [arXiv:hep-ex/0504003].

\bibitem{d0excl}
 T.~E.~Coan {\it et al.}  [CLEO Collaboration],
  Phys.\ Rev.\ Lett.\  {\bf 95}, 181802 (2005)
  [arXiv:hep-ex/0506052].

\bibitem{photos}
E. Barberio and Z. Was, Comput. Phys. Commun. {\bf 79}, 291(1994).

\bibitem{isgw}
  N.~Isgur, D.~Scora, B.~Grinstein and M.~B.~Wise,
  Phys.\ Rev.\ D {\bf 39}, 799 (1989).

\bibitem{cleo:dplus}
G.~S.~Huang {\it et al.}  [CLEO Collaboration],
  Phys.\ Rev.\ Lett.\  {\bf 95}, 181801 (2005)
  [arXiv:hep-ex/0506053].

\end{thebibliography}
\end{document}